\documentclass[runningheads]{llncs}
\usepackage[T1]{fontenc}
\usepackage{graphicx}
\usepackage{hyperref}
\hypersetup{pdftitle={Logic is Algebra},pdfauthor={Steven Obua}}
\usepackage{bookmark}
\usepackage{enumitem}
\usepackage{amssymb}
\usepackage{amsmath}
\usepackage{tgpagella}
\usepackage{setspace}
\usepackage{mathrsfs}
\usepackage{mathtools}
\usepackage{stmaryrd}

\begin{document}

\title{Logic is Algebra}
%
%
\author{Steven Obua\orcidID{0000-0002-4362-752X}}
\authorrunning{S. Obua}
%
\institute{Practal\\
\email{obua@practal.com}\\
\url{https://practal.com}}
\maketitle              
\begin{abstract}
Logic really is just algebra, given one uses the right kind of algebra, and the right kind of logic. 
The right kind of algebra is \emph{abstraction algebra}, and the right kind of logic is \emph{abstraction logic}. 

\end{abstract}

\newcommand{\minisection}[1]{\vspace{0.2cm}\noindent\textbf{#1.}}

\newcommand{\vars}{\mathcal{X}}
\newcommand{\terms}{\mathcal{T}}

\newcommand{\true}{\textsf{T}}
\newcommand{\false}{\textsf{F}}
\newcommand{\nats}{\mathbb{N}}
\newcommand{\fail}{\Finv}
\newcommand{\error}{\bot}
\newcommand{\imp}{\Rightarrow}
\newcommand{\all}{\forall}
\newcommand{\ex}{\exists}
\newcommand{\exu}{\exists_1}
\newcommand{\techterm}[1]{\textbf{#1}}
\newcommand{\deftechterm}[1]{\textbf{\textit{#1}}}
\newcommand{\univ}{\mathcal{U}}
\newcommand{\size}[1]{{|#1|}}
\newcommand{\sig}{{\mathfrak{S}}}
\newcommand{\logic}{{\mathcal{L}}}
\newcommand{\val}[1]{\emph{#1}}
\newcommand{\ctrue}{\true}
\newcommand{\cfalse}{\false}
\newcommand{\cimplies}{\imp}
\newcommand{\cforall}{\all}
\newcommand{\cexists}{\exists}
\newcommand{\eq}{=}
\newcommand{\ceq}{\eq}
\newcommand{\cnot}{\neg}
\newcommand{\cor}{\vee}
\newcommand{\cand}{\wedge}
\newcommand{\cequiv}{\Leftrightarrow}
\newcommand{\absatz}{\vspace{0.1cm}\noindent}
\newcommand{\valueof}[2]{\llbracket #2 \rrbracket_{#1}}
\newcommand{\appsubst}[2]{#2 / #1}
\newcommand{\valuationspace}{\mathcal{V}}
\newcommand{\algebra}{\mathcal{A}}
\newcommand{\valid}{\models}
\newcommand{\derives}{\vdash}
\newcommand{\pSUBST}{\textsc{Subst}}
\newcommand{\pINFER}{\textsc{Infer}}
\newcommand{\pTRUE}{\textsc{True}}
\newcommand{\timplies}{\text{$\imp$}}
\newcommand{\tall}{\text{$\all$}}
\newcommand{\ralg}{\mathfrak{R}}
\newcommand{\rclass}[1]{{[#1]_{\ralg}}}
\newcommand{\lax}[2]{$\text{#1}_{#2}$}

\newcommand{\operators}[1]{{\mathcal{O}(#1)}}
\newcommand{\carrier}[1]{{\mathcal{C}(#1)}}

\newcommand{\emphabs}[1]{``$\operatorname{#1}$''}

\section{The Right Logic}
What is the right logic? 
Obviously, the answer to this question depends on your specific circumstances. 
But if you are looking for a general logic as a foundation of mathematics and proof assistants,
then I have a proposition for you:
\begin{quote}
    The right logic is \emph{abstraction logic}. 
\end{quote}
At first, this statement sounds preposterous, 
especially when I tell you next that abstraction logic (AL) is simpler than any other general logic used in proof assistants,
such as first-order logic and simple or dependent type theory. 
If it existed, surely this ``right'' logic would have been discovered long ago?

I don't know why AL has been discovered only now. 
Maybe it actually \emph{has} been discovered before, and I just don't know about it. 
Given the vast literature on logic, this is possible. 
If so, it does not change much about the reason for this paper: \emph{Everyone} should be aware of 
abstraction logic and be using it.

To guess why AL has not been discovered before, 
it might help to reflect on how it \emph{was} discovered: 
I was looking for the right foundation for my proof assistant project \emph{Practal}~\cite{Practal}. Simple 
type theory felt too restrictive, dependent type theory too arbitrary, and first-order logic
could not even properly express binding constructs besides $\forall$ and $\exists$. This last observation
turned out to be essential\@: AL is similar to first-order logic, 
but instead of talking just about elements of a mathematical universe $\univ$, and functions and relations defined 
on $\univ$, AL allows talking about \emph{operators} on $\univ$. An operator is a functional on $\univ$, i.e.
a function which takes functions on $\univ$ as its input, and returns an element of $\univ$ as its output. 
Universal quantification $\forall$ and existential quantification $\exists$ are then just prominent operators
among many others. Elements of $\univ$, and functions and relations on $\univ$, are special cases of operators.

Of course, the idea that talking about functionals is useful is far from new. In mathematics, there is the whole branch 
of functional analysis based on just this insight. 
But instead of considering the entire mathematical universe $\univ$ as the topic under examination,
functional analysis typically focuses on special (albeit also rather general) mathematical structures such as vector spaces, Hilbert spaces, and topological spaces. 
More closely related to logic, mathematicians developed category theory, and computer scientists
developed type theory. And thus here is my guess why AL is not well-known today despite its simplicity: 
\begin{quote}
    AL got lost in the maelstrom of category theory and type theory.
\end{quote} 

The rest of this paper describes AL, and makes the case for AL being the right logic. 
First abstraction algebras are introduced as a generalization of abstract algebras. It is demonstrated
that abstraction algebras are expressive enough to straightforwardly model common mathematical concepts such as 
natural numbers, sets, functions, and simple and dependent types, even within a single abstraction algebra. 
Then AL's exceptionally simple syntax, semantics and proof theory are presented. AL is shown to be sound,
and also complete assuming modest conditions on the particular logic under consideration.
Finally, it is considered what it means for an abstraction logic to be inconsistent.
I conclude by briefly touching upon related and future work.

\section{Abstraction Algebra}
Let $A$ be a non-empty set\footnote{The word \emph{set} is used in a naive sense here.}. 
Every mapping $A^n \rightarrow A$ will be called an $n$-argument \deftechterm{operation} on $A$, 
or an operation of \deftechterm{arity} $n$. 
The notation $A_n$ serves as an abbreviation for $A^n \rightarrow A$. 
A zero-argument operation is interpreted as a constant element of $A$ and called a \deftechterm{value}. 
Therefore, $A_0$ and $A$ are identified.

Every mapping $A_{m_1} \times \dots \times A_{m_{n}} \rightarrow A$ is called an $n$-argument \deftechterm{operator},
or an operator of arity $n$. The list $[m_1, \ldots, m_n]$ is called its \deftechterm{(operator) shape}.
Because $A_0$ and $A$ are identified, every operation of arity $n$
is also an operator of arity $n$ with a shape consisting of zeros only. A value has shape $[\,]$.
An operation which is not a value, and an operator which is not an operation, are called 
\deftechterm{proper}.

An \deftechterm{abstraction algebra} $\algebra$ is a non-empty set $\carrier \algebra$, the \deftechterm{carrier of 
$\algebra$}, together with a set $\operators{\algebra}$ of operators on $\carrier \algebra$, the \deftechterm{operators of $\algebra$}. 

The notation $(A, o_1, \ldots, o_k)$ 
describes an abstraction algebra $\algebra$ with carrier $\carrier \algebra = A$ and operators 
$\operators{\algebra} = \{o_1, \ldots, o_k\}$.

\newcommand{\peanoalgebra}{\ensuremath{\textsl{Peano}}}
\newcommand{\funsetsalgebra}{\ensuremath{\textsl{FunSets}}}
\newcommand{\setsalgebra}{\ensuremath{\textsl{Sets}}}

\begin{example}
Any \techterm{(abstract) algebra}~\cite[p. 287]{Rasiowa-Modern-Mathematics} is an abstraction algebra 
$\algebra$ where every operator of $\algebra$ is an operation. Examples are the algebra of natural numbers
$(\nats, 0, 1, +, \cdot)$ and the two-element Boolean algebra $(\{\true, \false\}, \true, \false, \wedge, \vee, \neg)$.
\end{example}
The next three subsections present further examples: $\peanoalgebra$, $\setsalgebra$ and $\funsetsalgebra$. 

\newcommand{\opnat}{\textsl{N}}
\newcommand{\opnatapp}[1]{{\opnat(#1)}}
\newcommand{\opsuc}{\textsl{S}}
\newcommand{\opsucapp}[1]{{\opsuc(#1)}}
\newcommand{\opequals}{\equiv}
\newcommand{\opequalsapp}[2]{{#1\, \operatorname{\opequals}\, #2}}
\newcommand{\opimplies}{\Rightarrow}
\newcommand{\opimpliesapp}[2]{{#1\, \operatorname{\opimplies}\, #2}}
\newcommand{\opnot}{\neg}
\newcommand{\opnotapp}[1]{{\opnot\,#1}}
\newcommand{\opforall}{\forall}
\newcommand{\opforallapp}[2]{{\opforall\, (#1 \mapsto #2)}}
\newcommand{\opforallN}{\forall_\opnat}
\newcommand{\opforallNapp}[2]{{\opforallN\, (#1 \mapsto #2)}}

\subsection{$\peanoalgebra$}

Let $\peanoalgebra$ be the abstraction algebra
$$(\nats \uplus \{\true, \false, \error\}, \opnat, 0, \opsuc, \opequals, \opimplies, \opnot, \opforall, \opforallN).$$
where $\true$, $\false$ and $\error$ are three distinct values. The operator $0$ is a value and stands for itself. 
The other operators of $\peanoalgebra$ are defined as: 
\[
\begin{array}{lcl}
    \opnatapp x & \hspace{0.3cm}=\hspace{0.3cm} & \begin{cases} \true & \text{if $x \in \nats$} \\ \false & \text{otherwise} \end{cases} \\[0.5cm]
    \opsucapp x & = & \begin{cases} x + 1 & \text{if $x \in \nats$} \\ \error & \text{otherwise} \end{cases}\\[0.5cm]
    \opequalsapp x y & = & \begin{cases} \true & \text{if $x = y$} \\ \false & \text{otherwise} \end{cases} \\[0.5cm]
    \opimpliesapp x y \hspace{1cm}& = & \begin{cases} \false & \text{if $x = \true$ and $y \neq \true$} \\ \true & \text{otherwise} \end{cases} \\[0.5cm]
    \opnotapp x & = & \opimpliesapp{x}{\false} \\[0.1cm]
    \opforall\,f & = & \begin{cases} \true & \text{if $f(x) = \true$ for all $x \in \carrier \peanoalgebra$} \\ \false & \text{otherwise} \end{cases}\\[0.5cm]
    \opforallN\,f & = & \opforallapp x {\opimpliesapp {\opnatapp x} {f(x)}}
\end{array}
\]
Note that the following expressions
\begin{enumerate}\setstretch{1.2}
    \item $\opnatapp 0$,
    \item $\opforallNapp a {\opequalsapp a a}$,
    \item $\opforallNapp a {\opforallNapp b {\opimpliesapp {(\opequalsapp a b)} {(\opequalsapp b a)}}}$,
    \item $\opforallNapp a {\opforallNapp b {\opforallNapp c {\opimpliesapp{(\opequalsapp a b)}{(\opimpliesapp{(\opequalsapp b c)}{(\opequalsapp a c)})}}}}$,
    \item $\opforallapp a {\opforallNapp b {\opimpliesapp {(\opequalsapp a b)} {\opnatapp a}}}$,
    \item $\opforallNapp a {\opnatapp {\opsucapp a}}$,
    \item $\opforallNapp a {\opforallNapp b {\opimpliesapp {\opequalsapp {\opsucapp{a}} {\opsucapp{b}}} {\opequalsapp a b}}}$,
    \item $\opforallNapp a {\opnotapp {(\opequalsapp {\opsucapp a} 0)}}$,
    \item $\opimpliesapp {K(0)} {(\opimpliesapp {(\opforallNapp x 
        {\opimpliesapp {K(x)} {K(\opsucapp x)}})} 
        {\opforallN\,K})}$ 
\end{enumerate}
which correspond to the original Peano axioms~\cite[p. 113]{Peano-Axioms}, all evaluate to $\true$. 
In the last expression the 1-argument operation $K$ may be chosen arbitrarily.

\newcommand{\opfun}{\textsl{F}}
\newcommand{\opfunapp}[1]{\opfun(#1)}
\newcommand{\opdom}{\textsl{D}}
\newcommand{\opdomapp}[2]{\opdom(#1, #2)}
\newcommand{\opif}{\operatorname{if-then-else}}
\newcommand{\opifapp}[3]{\operatorname{if} #1 \operatorname{then} #2 \operatorname{else} #3}
\newcommand{\optag}[1]{\textsf{#1}}

\newcommand{\opin}{:}
\newcommand{\opsum}{\bigcup}
\newcommand{\opsumapp}[1]{\operatorname{\opsum} #1}
\newcommand{\opinterapp}[1]{\operatorname{\bigcap} #1}
\newcommand{\opreplace}{\mathscr{R}}
\newcommand{\opreplaceapp}[2]{\operatorname{\opreplace} #1\, #2}
\newcommand{\oppower}{\mathscr{P}}
\newcommand{\oppowerapp}[1]{\operatorname{\oppower} #1}
\newcommand{\opchoice}{\operatorname{\epsilon}}
\newcommand{\opchoiceapp}[1]{\opchoice #1}
\newcommand{\opthe}{\iota}
\newcommand{\optheapp}[1]{\operatorname{\opthe} #1}

\subsection{\setsalgebra}

Let $\setsalgebra$ be the abstraction algebra 
\[
    (\textsc{Sets}_{\true, \false}\ \uplus\ \{\error \}, 
    \opequals, \opif, \opforall, 
    \emptyset, \opin, \opsum, \oppower, \opreplace, \opthe, \opchoice, \opsuc, \nats).
\]
The carrier is the class of sets with \techterm{urelements} $\true$ and $\false$, and
a separate exceptional value $\error$. 
An urelement is just a value which is itself not a set, but can be an element of a set. 
The value $\error$ is therefore neither a set nor an urelement. 
The operation $\opequals$ is defined such that $(\opequalsapp A B) = \true$ if $A$ and $B$ are equal,
and otherwise $(\opequalsapp A B) = \false$. Two sets are equal if they have the same elements;
$\true, \false$ and $\error$ are distinct values unequal to any set and equal to only themselves.
The $\forall$ operator is also defined just as it was for $\peanoalgebra$, but of course taking into
account the different carrier. The conditional operation $\opif$ is defined via
\[
    (\opifapp C a b) = 
    \begin{cases}
        a & \text{if $C = \true$}, \\
        b & \text{otherwise}.
    \end{cases}
\]
The value $\emptyset$ is the empty set. The $:$ operation is defined as $(x : D) = \true$ if $D$ is
a set and $x \in D$, and as $(x : D) = \false$ otherwise. The $\opsum$ operation takes a set $D$ as its argument,
and returns the union of all sets $A \in D$. If $D$ is not a set, then $(\opsumapp D) = \emptyset$.
Similarly, if $D$ is a set, then $\oppowerapp D$ is the set of all subsets of $D$, and 
$(\oppowerapp D) = \emptyset$ if $D$ is not a set. The replacement operator $\opreplace$ takes a set $D$
and a unary operation $f$, replacing each element $x$ of $D$ by $f(x)$, yielding another set:
\[
    (\opreplaceapp D f) = \{\ y \mid \text{there is $x \in D$ such that $f(x) = y \neq \error$ } \}
\] 
If $D$ is not a set this implies $(\opreplaceapp D f) = \emptyset$. 
The definite description operator $\opthe$ selects the unique value for which an operation yields $\true$:
\[
    \optheapp f = 
    \begin{cases}
        x & \text{if there is a unique $x \in \carrier{\setsalgebra}$ such that $f(x) = \true$}\\
        \error & \text{otherwise}
    \end{cases}
\]
The choice operation $\opchoice$ picks an arbitrary element of a non-empty set:
\[
    \opchoiceapp D = 
    \begin{cases}
        x & \text{where $x$ is \emph{some} value such that $x \in D$}\\
        \error & \text{if $D$ is not a set, or if $D$ is an empty set}
    \end{cases}
\]
There is no further requirement on \emph{which} element $x$ of $D$ is picked, except of course that
$D = E$ implies $(\opchoiceapp D) = (\opchoiceapp E)$. Note that combining replacement with 
choice via $D \mapsto \opreplaceapp D \opchoice$ allows to satisfy \emph{the Axiom of Choice}. 

The set $\nats$ of natural numbers is instrumental in constructing infinite sets:
\[
    \nats = \left\{ \emptyset, \opsucapp{\emptyset}, \opsucapp{\opsucapp{\emptyset}}, 
    \opsucapp{\opsucapp{\opsucapp{\emptyset}}}, 
    \ldots \right\}.
\]
Here $\opsuc$ is the successor operation and defined as $\opsucapp n = n \cup \{n\}$.

There is a certain amount of leeway in how to define the operators of $\setsalgebra$. In the end, this is dependent
on the application you have in mind for your abstraction algebra. For this version of $\setsalgebra$,
the design principle has been to introduce fewer special cases instead of catching more potential 
specification errors. A reversal of this design principle would yield, for example, this definition instead:
\[
    (\opifapp C a b) = 
    \begin{cases}
        a & \text{if $C = \true$}, \\
        b & \text{if $C = \false$}, \\
        \error & \text{otherwise}.
    \end{cases}
\]

\subsection{\funsetsalgebra}

\newcommand{\opnotequalsapp}[2]{{#1 \not\equiv #2}}
\newcommand{\opandapp}[2]{{#1 \wedge #2}}
\newcommand{\oporapp}[2]{{#1 \vee #2}}
\newcommand{\opequivapp}[2]{{#1 \Leftrightarrow #2}}
\newcommand{\opupairapp}[2]{{\left\{ #1, #2 \right\}}}
\newcommand{\opsingleton}[1]{{\left\{ #1 \right\}}}
\newcommand{\opunionapp}[2]{{#1 \cup #2}}
\newcommand{\opintersectionapp}[2]{{#1 \cap #2}}
\newcommand{\opsepapp}[2]{\operatorname{\mathscr{S}} #1\, #2}
\newcommand{\bools}{\mathbb{B}}
\newcommand{\opfst}{\operatorname{first}}
\newcommand{\opfstapp}[1]{\opfst #1}
\newcommand{\opsnd}{\operatorname{second}}
\newcommand{\opsndapp}[1]{\opsnd #1}
\newcommand{\pfuns}[2]{#1 \rightharpoonup #2}
\newcommand{\funs}[2]{{#1 \rightarrow #2}}
\newcommand{\opapp}[2]{{#1 \cdot #2}}
\newcommand{\oplambda}[2]{{\operatorname{\lambda} {#1}\, #2}} 
\newcommand{\opdomain}[1]{{\operatorname{domain} #1}}
\newcommand{\opcodomain}[1]{{\operatorname{codomain} #1}}
\newcommand{\opfunctional}{\operatorname{functional}}
\newcommand{\opfunctionalapp}[1]{{\opfunctional #1}}
\newcommand{\dpairs}[2]{{\sum #1\,#2}}
\newcommand{\dfuns}[2]{{\prod #1\,#2}}

The abstraction algebra $\setsalgebra$ defines just a core of operators. More operators can be defined
based solely on this core. We do this until we reach $\funsetsalgebra$, an abstraction algebra
with the same carrier as $\setsalgebra$, but with the following additional operators, successively defined in terms of 
previously defined operators:
\[
\begin{array}{lcl}
    \true &\hspace{0.3cm} = \hspace{0.3cm}& \opforallapp x {\opequalsapp x x}\\
    \false & = & \opforallapp x x \\
    \error & = & \optheapp {(x \mapsto \false)} \\
    \opimpliesapp A B & = & \opifapp A B \true\\
    \opnotapp A & = & \opimpliesapp A \false \\
    \opnotequalsapp A B & = & \opnotapp{(\opequalsapp A B)} \\
    \opandapp A B & = & \opifapp A B A \\
    \oporapp A B & = & \opifapp A A B \\
    \opequivapp A B & = & \opandapp {(\opimpliesapp A B)} {(\opimpliesapp B A)}\\
    \opupairapp a b & = & \opreplaceapp {(\oppowerapp {(\oppowerapp \emptyset)})} 
        {(x \mapsto \opifapp {\opequalsapp x \emptyset} a b)} \\
    \opsingleton a & = & \{ a, a \} \\
    \opunionapp D E & = & \opsumapp {\opupairapp D E}\\
    D \subseteq E & = & \opforallapp x {\opimpliesapp {(x : D)} {(x : E)}}\\
    \bools & = & \{ \true, \false \}\\    
    \opsepapp D p & = & \opreplaceapp D {(x \mapsto \opifapp {p(x)} x \error)} \\
    \opinterapp M & = & \opsepapp {(\opsumapp M)} {(x \mapsto \opforallapp D {\opimpliesapp {(D : M)} {(x : D)}})} \\
    \opintersectionapp D E & = & \opsepapp D {(x \mapsto x : E)}\\
    \exists\, p & = & \opnotapp {\opforallapp x {\opnotapp{p(x)}}}\\
    (a, b) & = & \opifapp {\opandapp {\opnotequalsapp a \error} {\opnotequalsapp b \error}} {\opupairapp a {\opupairapp a b}} \error\\
    \opfstapp u & = & \optheapp {(x \mapsto \exists\, {(y \mapsto \opequalsapp u {(x, y)})})}\\
    \opsndapp u & = & \optheapp {(y \mapsto \exists\, {(x \mapsto \opequalsapp u {(x, y)})})}\\
    D \times E & = & \opsumapp {\opreplaceapp D { (d \mapsto \opreplaceapp E { (e \mapsto (d, e))})}}\\
    \opdomain R & = & \opreplaceapp R \opfst \\
    \opcodomain R & = & \opreplaceapp R \opsnd \\
    \opfunctionalapp R & = & \opforallapp x {\opforallapp a {\opforallapp b {\opimpliesapp 
        {(\opandapp {(x, a) : R} {(x, b) : R})}{\opequalsapp a b}}}}\\
    \pfuns D E & = & \opsepapp {(\oppowerapp {(D \times E)})} \opfunctional \\
    \funs D E & = & \opsepapp {(\pfuns D E)} { (R \mapsto \opequalsapp {\opdomain R} D) }    \\
    \opapp R x & = & \optheapp {(y \mapsto (x, y) : R)}\\
    \oplambda D f & = & \opreplaceapp D {(x \mapsto (x, f(x)))}\\
    \dpairs D t & = & \opsepapp {(D \times {\opsumapp {\opreplaceapp D t}})} 
        {(u \mapsto (\opsndapp u) : t(\opfstapp u)) } \\
    \dfuns D t & = & \opsepapp {(\funs D {\opsumapp {\opreplaceapp D t}})} 
        {(f \mapsto \opforallapp x {\opimpliesapp {(x : D)} {(f(x) : t(x))}})} \\
\end{array}
\]
With these definitions, the following example expressions are all equal to $\true$,
for any instantiation of free variables appearing in the respective expression:
\begin{itemize}
    \item $\opnotequalsapp \false \error$, but also $\opequivapp \false \error$,
    \item $\opequalsapp {\{a, \error\}} {\{a\}}$, $\opequalsapp {\{\error, a\}} {\{a\}}$,
        $\opequalsapp {\{ \error \}} \emptyset$, $\opequalsapp {(a, \error)} \error$, $\opequalsapp {(\error, a)} \error$,
    \item $\opequalsapp {\left(\opsumapp \nats\right)} \nats$, and $\opequalsapp {\left(\opsumapp \bools\right)} \emptyset$,
    \item $\opequalsapp {\dpairs D {(x \mapsto E)}} {D \times E}$, $\opequalsapp {\dfuns D {(x \mapsto E)}} {\funs D E}$,
    \item $\opimpliesapp {\opandapp {(x : D)} {(F : \funs D E)}} {(\opapp F x) : E}$,
    \item $\opimpliesapp {(x : D)} {\opequalsapp 
      {\opapp {(\oplambda D f)} x} {f(x)}}$.
\end{itemize}

\subsection{Data Abstraction}
A common complaint about set theory is that it makes \techterm{data abstraction} impossible, 
because everything is a set. This is less of a problem with $\setsalgebra$ and $\funsetsalgebra$,
because one could just ``forget'' the definition of an object, and work with its abstract properties
instead. For example, one could forget how the elements of $\nats$ have been defined, and instead
just work with them as one would within $\peanoalgebra$. This way it is impossible to know
if a natural number is a set, because it could always be an urelement! In particular, the equation
$\bigcup \nats = \nats$, which is true in $\setsalgebra$, is then not known to hold. 
But this uncertainty might not be enough for certain applications. 
It might be required to be \emph{certain} that the elements of $\nats$ are \emph{not sets},
so that $\bigcup \nats = \emptyset$ holds, just as $\bigcup \bools = \emptyset$ already holds.

Proof assistants based on simple type theory such as HOL4, Hol Light, and Isabelle/HOL, feature
a type definition mechanism to introduce a new type based on a bijection relating it to a subset of an existing type. 
The same principle can be applied to $\funsetsalgebra$. First, the existence of 
an ordered pair operation $x, y \mapsto \langle x, y \rangle$ is postulated, 
such that $\langle x, y \rangle$ is guaranteed to be an urelement. Such a construction
has been done by Dunne et al.\ in~\cite{Abstraction-Barrier} to create an \emph{abstraction barrier} on top of 
ZF set theory\footnote{See also my MathOverflow question 
\url{https://mathoverflow.net/questions/366014/data-abstraction-in-set-theory-via-urelements}.}. 
Then, to create a new data type based on an existing set $S$, 
a new \deftechterm{tag} $\tau$ is created which is guaranteed to be different from previously created
tags, and the set $T = \{ \langle \tau, s \rangle \mid s \in S \}$ constitutes the new data type $T$.
This way, tags $\texttt{bool}$, $\texttt{nat}$, $\texttt{pfun}$, $\texttt{fun}$, $\texttt{dpair}$
and $\texttt{dfun}$ can be introduced to create abstract data types not only for booleans $\bools$,
but also for natural numbers $\nats$, function types $\pfuns D E$ and $\funs D E$, and 
dependent types $\dpairs D t$ and $\dfuns D t$.

\section{Just Use a Lambda\texttrademark}

So far expressions have been formed from operators in a standard mathematical fashion. The task
now is to define \techterm{terms}, such that when a term is evaluated with respect to 
an abstraction algebra, it corresponds to such an expression. As an example, consider the definition
of $\lambda$ for $\funsetsalgebra$:
\[
    \oplambda D f \quad = \quad \opreplaceapp D {(x \mapsto (x, f(x)))}.
\] 
The defining expression consists of a symbol $\opreplace$ naming an operator taking two arguments: 
a value $D$, and a unary operation on $\carrier{\funsetsalgebra}$ represented by the expression
$x \mapsto (x, f(x))$. The temptation is to \emph{Just Use a Lambda}\texttrademark\ 
to define terms that can \emph{directly} represent such an operation, for example via
$\operatorname{\Lambda} x. (x, f(x))$.

The problem with this is that the carrier of an abstraction algebra corresponds to an
entire \deftechterm{mathematical universe}. This mathematical universe consists of \techterm{values}.
Introducing a Lambda is an attempt to treat operations as normal elements of the mathematical universe, 
just as values are. But due to Cantor's theorem there are many more operations than there are values
(assuming there are at least two values). This means we cannot just bunch operations together with values, 
and call them values 2.0, as that would lead to a paradox. Trying to make operations part of the 
mathematical universe therefore introduces the need for
\techterm{types}, in order to separate values from operations within the mathematical universe. 

\emph{But types don't buy you anything.} Types are an arbitrary restriction on your freedom to combine
operators in any way you choose to, and disallow many useful operators entirely. Trying to gain some of that freedom back after the introduction
of types then leads to ever more complex type systems, raging within 
Barendregt's lambda cube~\cite{Lambda-Cube}. But as we have known since the invention of set theory,
and as has again been demonstrated here concretely with $\funsetsalgebra$, sets, functions and types (as sets) \emph{can} live as one
within a single mathematical universe. \emph{It's just that not all operations can.} 
But certainly the entire lambda cube can fit into this mathematical universe,
and typing rules just become normal theorems. As it turns out, those theorems you thought you got for 
free by embracing types~\cite{Theorems-for-free}, they were never free. The price you paid was freedom itself.

The alternative to a type system is to not let operations exist on their own as terms. Of course, it
still needs to be possible to talk about operations, and that is what operators are for. 
Talking about operations is just applying operators to operations.
And so the solution is to define terms in such a way that operator application is represented directly,
without granting operations (or operators, for that matter) an existence outside such an application:
\[
    \lambda\,x.\, D\, f[x] \quad \opequals \quad \opreplace\,x.\, D\, (x, f[x]).
\]
This way, every term evaluates to a value. The next sections make this precise.

\section{Syntax}

\newcommand{\natsof}[1]{\overline{#1}}
\newcommand{\abs}[1]{{|#1|}}

A \deftechterm{signature} is a set of names called \deftechterm{abstractions}, where each
abstraction is associated with an \deftechterm{(abstraction) shape}. An abstraction shape is
a list $[p_1, \ldots, p_n]$ such that each $p_i$ is a subset of $\natsof{m} = \{1, 2, \ldots, m\}$,
and such that $\natsof{m} = \bigcup_{i = 1}^n p_i$. 
Then $n$ is called the \deftechterm{arity} of the shape / abstraction,
and $m$ is called its \deftechterm{valence}.  

A $\sig$-algebra is an abstraction algebra $\algebra$ together 
with a signature $\sig$ such that each abstraction $a \in \sig$ of shape $[p_1, \ldots, p_n]$ 
is associated with some operator $o \in \operators{\algebra}$ of shape $[\abs{p_1}, \ldots, \abs{p_n}]$.
We will usually not distinguish between the $\sig$-algebra and the abstraction algebra, and refer to
both of them by the same name, in this case $\algebra$. 

Assume an infinite set $X$ of \deftechterm{variables}. A \deftechterm{term}, relative to $X$ and a given signature
$\sig$, is either a variable application or an abstraction application.

A \deftechterm{variable application} has the form
\[
    x[t_1, \ldots, t_n],
\]
where $x \in X$ and the $t_i$ are themselves terms. We say that $x$ occurs with arity $n$. In case
of $n = 0$ we also just write $x$ instead of $x[\,]$.

An \deftechterm{abstraction application} has the form
\[
    (a\,x_1 \ldots x_m.\, t_1 \ldots t_n),
\]
where $a$ is an abstraction belonging to the signature, $m$ is the valence of $a$, and $n$ is the arity 
of $a$. The $t_i$ are terms representing the operations to which $a$ is being applied. 
The $x_j$ are distinct variables. The idea is that the shape $[p_1, \ldots, p_n]$ 
of the abstraction determines which of these variables are bound by each $t_i$. 
E.g., for $p_i = \{q_1 < \cdots < q_k\}$, the term $t_i$ binds the 
variables $x_{q_1}, \ldots, x_{q_k}$ and represents the $k$-ary operation 
$x_{q_1}, \ldots, x_{q_k} \mapsto t_i$. Here the notation $\{q_1 < \cdots < q_k\}$ is short
for $\{q_1, \ldots, q_k\}$, where $q_1 < \cdots < q_k$.

There is more than one way of how to choose the $p_i$ when representing an operator by an abstraction,
at least for proper $n$-ary operators with $n > 1$.
Assuming the operator has shape $[1, 1]$, the shape of the abstraction could be $[\{1\}, \{2\}]$ or 
[\{2\}, \{1\}], resulting in a valence of $2$, or it could be $[\{1\}, \{1\}]$, resulting in a valence of $1$.   
E.g, if the meaning of the operator is the addition of two arrays $u$ and $v$, then an abstraction of valence $1$
seems more appropriate, as the intuition is that addition is performed in lock-step: 
$(\operatorname{add-arrays} i.\, u[i]\, v[i])$.
On the other hand, if the meaning of the operator is to perform the addition such that each element
of $u$ is added to every element of $v$ such that the result forms a matrix, then independent indexing seems better:
$(\operatorname{matrix-add-arrays} i\, j.\, u[i]\, v[j])$.
But it would be perfectly fine, although maybe of questionable taste, to choose representations 
$(\operatorname{add-arrays} i\, j.\, u[i]\, v[j])$ and $(\operatorname{matrix-add-arrays} i.\, u[i]\, v[i])$ instead.

Instead of writing $(a\, x_1 \ldots x_m.\, t_1 \ldots t_n)$, for concrete cases 
obvious custom syntax may be used, such as binary operation syntax for $n=2$ and $m=0$. 

\section{Semantics}

Let $\algebra$ be a $\sig$-algebra, and $X$ a set of variables. 
A \deftechterm{valuation} $\nu$ into $\algebra$ assigns to each 
variable $x \in X$ and each arity $n$ an $n$-ary operation on $\carrier{\algebra}$. 

For distinct variables $x_1, \ldots, x_{k}$ and values $u_1, \ldots, u_{k}$ in $\carrier{\algebra}$, 
an updated valuation is defined by 
\[
    \nu[x_1 \coloneqq u_1, \ldots, x_{k} \coloneqq u_{k}](y, n) = 
    \begin{cases}
        u_i & \text{for}\ y = x_i\ \text{and}\ n = 0, \\ 
        \nu(y, n) & \text{otherwise}.
    \end{cases}
\]

A valuation $\nu$ gives meaning to all variables in $X$, and $\algebra$ gives meaning 
to all abstractions in $\sig$. This makes it possible to \deftechterm{calculate} the value $\valueof \nu t$ 
of any term $t$ with respect to $\nu$ and $\algebra$ by recursing over the structure of $t$:
\begin{itemize}
\item If $t = x[t_1, \ldots, t_n]$ is a variable application, then 
\[ \valueof \nu t = \nu(x, n)(\valueof \nu {t_1}, \ldots, \valueof \nu{t_n}). \]
\item If $t = (a\, x_1 \ldots x_m.\, t_1 \ldots t_n)$ is an abstraction application, then  
let $o$ be the $n$-ary operator associated with $a$, and let $[p_1, \ldots, p_n]$ be the shape of $a$.
For $p_i = \{q_1 < \cdots < q_k\}$, the $k$-ary operation $f_i$ on $\carrier{\algebra}$ is defined as 
\[ (u_1, \ldots, u_k) \mapsto \valueof {\nu[x_{q_1} \coloneqq u_1, \ldots, x_{q_k} \coloneqq u_k]} {t_i}, \]
from which we obtain $\valueof \nu t = o(f_1, \ldots, f_n)$. 
\end{itemize}
The identity of a variable is inextricably linked with the arity it occurs with. The term
$x[x]$ really means $x^1[x^0]$, with $x^1$ being a different variable than $x^0$.

Two terms $s$ and $t$ are called \deftechterm{semantically equivalent} with respect to a signature
$\sig$ if for all $\sig$-algebras $\algebra$ and all valuations $\nu$ into $\algebra$,
$\valueof \nu s = \valueof \nu t$ holds. 
Furthermore, two terms are called \deftechterm{$\alpha$-equivalent}
if their representations using \techterm{de Bruijn indices}~\cite{AL} are identical. 
It can be shown that two terms are semantically equivalent if and only if 
they are $\alpha$-equivalent.

A variable $x$ occurs \deftechterm{free} with arity $n$ in a term $t$ if there is an occurrence of 
$x$ with arity $n$ in $t$ not bound by any surrounding abstraction. Because abstractions 
bind only variables of arity $0$, \emph{any} occurrence of $x$ in $t$ with arity $n > 0$ is free.

It is clear that the value of $t$ depends only on the assignments in the valuation to those 
variables $x$ which are free in $t$. Therefore, the value of a \deftechterm{closed} term $t$, 
i.e.\ a term without any free variables, does not depend on the valuation at all, but only on the 
abstraction algebra in which the calculation takes place.

\newcommand{\peanosig}{\sig_{\peanoalgebra}}
\newcommand{\funsetssig}{\sig_{\funsetsalgebra}}
\begin{example}
    Consider the abstraction algebra $\peanoalgebra$. The signature $\peanosig$ is formed
    by reusing the operator names of $\peanoalgebra$ also as abstraction names. Then
    \[
        \opimpliesapp {K[0]} {(\opimpliesapp {(\opforallN\,x.\, \opimpliesapp {K[x]} {K[(\opsuc.\, x)]})} {(\opforallN\, x.\, K[x])})}
    \] 
    evaluates to $\true$ for any valuation into $\peanoalgebra$.
\end{example}
\begin{example}
    Consider the abstraction algebra $\funsetsalgebra$. The signature $\funsetssig$ is formed
    by reusing the operator names of $\funsetsalgebra$ as abstraction names. Then
    \[
        \begin{array}{c}
        \opequalsapp {\sum x.\, D\, E} {D \times E}, \\[0.1cm]
        \opequalsapp {\prod x.\, D\, E} {\funs D E},\\[0.1cm]
        \opimpliesapp {\opandapp {(x : D)} {(F : \funs D E)}} {(\opapp F x) : E},  \\[0.1cm]     
       \opimpliesapp {( x : D )} {\opequalsapp {\opapp {(\operatorname{\lambda} x.\, D\, f[x])} x} {f[x]}}
        \end{array}
    \] 
    all evaluate to $\true$ for any valuation into $\funsetsalgebra$.
\end{example}

\section{Substitution}

An \deftechterm{$n$-ary template} has the form $[x_1 \ldots x_{n}.\, t]$,
where the $x_i$ are distinct variables called \deftechterm{template binders} and $t$ is a term
called \deftechterm{template body}. 
The template binds all free occurrences of 
$x_i$ in $t$ of arity $0$. A valuation $\nu$ turns an $n$-ary template into an $n$-ary operation via 
\[ \valueof \nu {[x_1 \ldots x_{n}.\, t]} \hspace{0.3cm} = \hspace{0.3cm} (u_1, \ldots, u_n) \mapsto 
\valueof {\nu[x_1 \coloneqq u_1, \ldots, x_{n} \coloneqq u_{n}]} t. \]
It is straightforward to extend the notion of semantical and $\alpha$-equivalence from terms to templates. 
Two templates $U$ and $V$ are called \deftechterm{semantically equivalent}
with respect to a signature $\sig$ if for all $\sig$-algebras $\algebra$ and all 
valuations $\nu$ into $\algebra$, $\valueof \nu U$ and $\valueof \nu V$ are the same operations.
Two templates are called $\alpha$-equivalent if their representation using
\techterm{de Bruijn indices}~\cite{AL} is identical. As with terms,
semantical and $\alpha$-equivalence turn out to be the same.

A \deftechterm{substitution} $\sigma$ is a function defined on a \techterm{domain} $D$ that maps a 
variable $x$, given an arity $n$ such that $(x, n)$ belongs to $D$, to an $n$-ary template $\sigma(x, n)$.
The purpose of a substitution $\sigma$ is to be applied to a term/template $t$, yielding another term/template
$\appsubst{\sigma}{t}$ as the result of the substitution. In principle this is done by replacing free variables in 
$t$ which are in the domain of $\sigma$ by their corresponding templates, and then in turn applying these 
templates to their arguments again by substitution. The details of this are somewhat
intricate, as care must be taken to avoid the capture of free variables by bound variables. 
Maybe the easiest way to do this is to convert terms/templates to a representation based on de Bruijn indices,
perform the substitution, and convert the result back to normal terms/templates.
As a consequence, there is no canonical result of applying a substitution: 
The result is determined only up to $\alpha$-equivalence. 

The main property of substitution is that, given a background valuation $\nu$,
any substitution $\sigma$ can be turned into a valuation $\nu_\sigma$
via $\nu_\sigma(x, n) = \valueof \nu {\sigma(x, n)}$ for $(x, n) \in D$, and 
$\nu_\sigma(x, n) = \nu(x, n)$ for $(x, n) \notin D$.  
The valuation $\nu_\sigma$ has the property that for any term or template $t$ the following holds:
$\valueof{\nu_\sigma}{t} = \valueof \nu {\appsubst{\sigma}{t}}$.

\section{Logic, Truth, and Models}\label{logic}
\newcommand{\prems}{\mathit{P}}
\newcommand{\concl}{\mathit{c}}

It is straightforward to turn abstraction algebra into a logic. That this is possible
is maybe not surprising\footnote{It was surprising to me, though. 
Search for ``inconsistent'' in~\cite{PracticalTypes}.}, but certainly not self-evident. 
After all, Church tried exactly that with his untyped lambda calculus, and failed~\cite{Kleene-Rosser-Inconsistency}. 
Only with the introduction of types he was able to turn lambda calculus into a logic~\cite{Church-Simple-Types}. 
I guess Church was the first victim of \emph{Just Use a Lambda}\texttrademark.

A \deftechterm{premise} is a template, such that all of its template binders appear free in its
template body. Two premisses are identified if they differ in the order of their template binders only.
A \deftechterm{rule} is a pair $(\prems, \concl)$, where $\prems$ is a finite set of premisses, 
and $\concl$ is a term called \emph{conclusion}. If $\prems$ is empty, we also write just $\concl$.

An \deftechterm{abstraction logic} $\logic$ is a signature $\sig$ together with a set of 
rules called \deftechterm{inference rules}. An inference rule without premisses is called an \deftechterm{axiom}.
 
Let $\algebra$ be a $\sig$-algebra and let $\true \in \carrier \algebra$ be a designated value of the algebra.
A term $t$ is called \deftechterm{true} in $\algebra$ for $\nu$, if $\valueof \nu t = \true$. A template
$U$ is called \deftechterm{true} in $\algebra$ for $\nu$ if $\valueof \nu U$ is a
constant operation equalling $\true$ everywhere. A rule
is called \deftechterm{true} for $\nu$ in $\algebra$ if either its conclusion is true for $\nu$ in $\algebra$, 
or one of its premisses is \emph{not} true for $\nu$ in $\algebra$.

A \deftechterm{valuation space} $\valuationspace$ for $\algebra$ is a set of 
valuations into $\algebra$ such that:
\begin{itemize}
\item $\valuationspace$ is not empty. 
\item If $\nu$ is a valuation belonging to $\valuationspace$, $u \in \carrier \algebra$, 
  and $x$ is a variable, then $\nu[x \coloneqq u]$ also belongs to $\valuationspace$.
\item If $\nu$ is a valuation belonging to $\valuationspace$, and if $\sigma$ is a substitution (with 
respect to $\sig$), then $\nu_\sigma$ also belongs to $\valuationspace$.
\end{itemize}
This notion of valuation space turns out to be strong enough to prove \emph{soundness} of 
abstraction logic, and weak enough to treat its \emph{completeness}.\ Of course, 
the space of \emph{all} valuations into $\algebra$ is trivially a valuation space.

A \deftechterm{model} for $\logic$ is a triple $(\algebra, \true, \valuationspace)$ such that
$\algebra$ is a $\sig$-algebra, $\true \in \carrier \algebra$ a designated but arbitrary value 
denoting \deftechterm{truth}, $\valuationspace$ a valuation space for $\algebra$, and such that all
inference rules of $\logic$ are true in $\algebra$ for all $\nu \in \valuationspace$. 
The model is called \deftechterm{standard} if $\valuationspace$ is the space of all valuations
into $\algebra$, and otherwise \deftechterm{non-standard}.

A model is called \deftechterm{degenerate} if the carrier of $\algebra$ consists of a single value,
which necessarily is $\true$. 
Note that \emph{every} abstraction logic has degenerate models. 

If every model of $\logic$ is also a model of the logic with the same signature as $\logic$ and 
the single inference rule $r$, then we say that $r$ is \deftechterm{valid} in $\logic$ and write 
\[\logic \valid r.\]

\section{Proofs}\label{proofs}
\newcommand{\proofof}[1]{\operatorname{p}_#1}

Two rules are called $\alpha$-equivalent if their conclusions are $\alpha$-equivalent,
and for each premise of one rule there is an $\alpha$-equivalent premise of the other rule.
Application of a substitution $\sigma$ to a rule $r$ yields the rule $\appsubst \sigma r$ resulting
from applying $r$ to the premisses (and normalizing the resulting templates by removing unused template binders)  
and the conclusion of $r$.

Given a logic $\logic$ with signature $\sig$, 
a \deftechterm{proof} $\proofof r$ in $\logic$ of a rule $r$ is either a \techterm{truism}, 
a \techterm{substitution}, or an \techterm{inference}:
\begin{itemize}
\item A \deftechterm{truism} $\pTRUE(r)$ proves $r$, where $r$ is an inference rule.
\item A \deftechterm{substitution} $\pSUBST(r, \proofof s, \sigma)$ proves $r$, 
where $\proofof s$ is a proof of $s$, 
and $r$ is $\alpha$-equivalent to $\appsubst \sigma s$.
\item An \deftechterm{inference} $\pINFER(r, \proofof s, \proofof t)$ proves $r$, under the following assumptions:
\begin{itemize}
    \item $\proofof s$ is a proof of $s = (\{H_1, \ldots, H_n\}, c)$, where $H_1 = [x_1 \ldots x_k.\, h]$,  
    \item $\proofof t$ is a proof of $t = (\{G_1, \ldots, G_m\}, h)$, where none of
    the template binders of any $G_j$ equals any of the $x_i$, and
    \item $r = (\{G'_1, \ldots, G'_m, H_2, \ldots, H_n\}, c)$, where $G_j$ yields $G'_j$ 
    by adding those $x_i$ as template binders which occur free with arity 0 in the body of $G_j$.
\end{itemize}
\end{itemize}
If there exists a proof of rule $r$ in $\logic$, we say that $r$ is a \deftechterm{theorem} of $\logic$ and write 
\[ \logic \derives r.\]

\section{Soundness}\label{soundness}

Abstraction logic is \deftechterm{sound}, i.e.\ every theorem of $\logic$ is also valid in $\logic$:
\[\logic \derives r \quad\text{implies}\quad \logic \valid r \quad\text{for every rule $r$.}\]
We show by induction over the structure of proofs that if $\proofof r$ is a proof of $r$ in $\logic$, 
then $r$ is valid in $\logic$:
\begin{itemize}
\item Assume $\proofof r = \pTRUE(r)$. Then $r$ is an inference rule, and is therefore true in every model of $\logic$.
\item Assume $\proofof r = \pSUBST(r, \proofof s, \sigma)$, where $\proofof s$ is a proof of $s$, and
\[r = (\{H_1, \ldots, H_m\}, d) \quad\text{and}\quad s = (\{G_1, \ldots, G_n\}, c).\] 
Here $d$ and $\appsubst \sigma c$ are $\alpha$-equivalent and therefore also semantically equivalent, and
for each $i \in \natsof n$ there is a $j \in \natsof m$ such that $H_j$ and $\appsubst \sigma {G_i}$ 
are semantically equivalent, too.
Now let $(\algebra, \true, \valuationspace)$ be a model of $\logic$, and $\nu \in \valuationspace$. Then
\[
\valueof \nu d = \valueof \nu {\appsubst \sigma c} = \valueof {\nu_\sigma} {c},\quad\text{and}\quad
\valueof \nu {H_j} = \valueof \nu {\appsubst \sigma {G_i}} = \valueof {\nu_\sigma} {G_i}.
\]
Because $\valuationspace$ forms a valuation space, 
$\nu_\sigma \in \valuationspace$, and because $s$ is valid, either 
$\valueof {\nu_\sigma} {c} = \true$ or there is $i \in \natsof n$ such that 
$\valueof {\nu_\sigma} {G_i}(u_1, \ldots, u_k) \neq \true$ for some $u_1, \ldots, u_k \in \carrier \algebra$. 
But this implies that either
$\valueof \nu d = \true$, or there is $j \in \natsof m$ such that 
$\valueof \nu {H_j} (u_1, \ldots, u_k) \neq \true$. Thus, $r$ is valid.
\item Assume $\proofof r = \pINFER(r, \proofof s, \proofof t)$, where 
$s = (\{H_1, \ldots, H_n\}, c)$, 
$H_1 = [x_1 \ldots x_k.\, h]$, 
$t = (\{G_1, \ldots, G_m\}, h)$, and
$r = (\{G'_1, \ldots, G'_m, H_2, \ldots, H_n\}, c)$.
Let $(\algebra, \true, \valuationspace)$ be a model of $\logic$, and $\nu \in \valuationspace$.
We need to show that $r$ is true for $\nu$, assuming that both $s$ and $t$ are valid in $\logic$.
If $\valueof \nu c = \true$, then $r$ is true for $\nu$, and we are done. 
So assume $\valueof \nu c \neq \true$. Because of the validity of $s$, 
for some $i \in \natsof n$, $H_i$ is not true with respect to $\nu$. If $i \geq 2$, 
then $H_i$ is also a premise of $r$; thus $r$ is true and we are done again. 
So assume $i = 1$. Then for some $u_1, \ldots, u_k \in \carrier \algebra$, and 
$\nu' = \nu[x_1 \coloneqq u_1, \ldots, x_k \coloneqq u_k]$, 
we have $\valueof {\nu'} h \neq \true$. Because $\nu' \in \valuationspace$, and $t$ is valid,
this implies that there is $j\in \natsof m$ such that $\valueof {\nu'} {G_j} \neq \true$.
For $G_j = [y_1 \ldots y_l.\, g]$ with $\{y_1, \ldots, y_l\} \cap \{x_1, \ldots, x_k\} = \emptyset$, 
this means that there are $v_1, \ldots, v_l \in \carrier \algebra$
such that for $\nu'' = \nu'[y_1 \coloneqq v_1, \ldots, y_l \coloneqq v_l]$, we have
$\valueof {\nu''} g \neq \true$. Let $x_{\alpha_1}, \ldots, x_{\alpha_q}$ 
for $\{\alpha_1 < \cdots < \alpha_q\} \subseteq \natsof k$ be those variables among
the $x_1, \ldots, x_k$ which appear free in $g$ with arity 0, and assume 
$G'_j = [x_{\alpha_1} \ldots x_{\alpha_q}\, y_1 \ldots y_l.\, g]$. Then
$\valueof \nu {G'_j}(u_{\alpha_1}, \ldots, u_{\alpha_q}, v_1, \ldots, v_l) =
\valueof \nu {[x_{\alpha_1} \ldots x_{\alpha_q}\, y_1 \ldots y_l.\, g]}(u_{\alpha_1}, \ldots, u_{\alpha_q}, v_1, \ldots, v_l) =
\valueof {\nu''} g \neq \true$. Therefore, $G'_j$ is not true with respect to $\nu$, 
which makes $r$ true for $\nu$.
\end{itemize}

\newcommand{\delogic}{{\logic_E}}
\section{Deduction Logic with Equality}
An important abstraction logic is \deftechterm{deduction logic with equality} $\delogic$. Its signature
consists of four abstractions: truth $\true$, implication $\opimplies$, equality $\opequals$,
and universal quantification $\opforall$. It has the following 10 inference rules, 8 of which are axioms:
\begin{center}
\begin{tabular}{r@{\hskip 0.5cm}l}
    \textsc{Modus Ponens} & $(\{\opimpliesapp A B,\ A \}, B)$ \\
    \textsc{Universal Introduction} & $(\{[x.\ P[x]]\}, \forall x.\, P[x])$ \\
    $\textsc{Truth}_1$ & $\true$ \\
    $\textsc{Truth}_2$ & $\opimpliesapp A {(\opequalsapp A \true)}$ \\
    $\textsc{Implication}_1$ & $\opimpliesapp A {(\opimpliesapp B A)}$ \\
    $\textsc{Implication}_2$ & 
        $\opimpliesapp {(\opimpliesapp A {(\opimpliesapp B C)})} 
        {(\opimpliesapp {(\opimpliesapp A B)} {(\opimpliesapp A C)})}$ \\
    $\textsc{Universal}_1$ & $\opimpliesapp {(\forall x.\, A[x])} {A[x]}$ \\
    $\textsc{Universal}_2$ & $\opimpliesapp {(\forall x.\, \opimpliesapp A {B[x]})} 
       {(\opimpliesapp A {(\forall x.\, B[x])})}$\\
    $\textsc{Equality}_1$ & $\opequalsapp x x$\\
    $\textsc{Equality}_2$ & $\opimpliesapp {(\opequalsapp x y)} {(\opimpliesapp {A[x]} {A[y]})}$
\end{tabular}
\end{center}
A logic $\logic'$ is called an \deftechterm{axiomatic extension} of a logic $\logic$ if every abstraction
of $\logic$ is also an abstraction of $\logic'$ of the same shape, if for every inference rule
of $\logic$ there is a corresponding $\alpha$-equivalent inference rule of $\logic'$,
and if every inference rule of $\logic'$ which has no corresponding inference rule of $\logic$ is
actually an axiom.
It is straightforward to construct logics $\logic_\peanoalgebra$ and $\logic_\setsalgebra$ which are
axiomatic extensions of $\delogic$ such that $(\peanoalgebra, \true, \valuationspace_\peanoalgebra)$ is
a model of $\logic_\peanoalgebra$, and $(\setsalgebra, \true, \valuationspace_\setsalgebra)$ is 
a model of $\logic_\setsalgebra$. Because $\funsetsalgebra$ is a definitional extension
of $\setsalgebra$, $\logic_\funsetsalgebra$ is constructed by adding those definitions as axioms
to $\logic_\setsalgebra$, and $(\funsetsalgebra, \true, \valuationspace_\funsetsalgebra)$ is a model.
Here $\valuationspace_\peanoalgebra$, $\valuationspace_\setsalgebra$, and $\valuationspace_\funsetsalgebra$
consist of all valuations into the respective abstraction algebras.

\newcommand{\termeq}{\approx}
\newcommand{\termeqapp}[2]{#1 \termeq{} #2}
\newcommand{\classof}[1]{[#1]}
\newcommand{\termclasses}{\mathcal{U}}
\newcommand{\rasiowa}{\mathfrak{R}}
\newcommand{\templatefuncsymbol}{\Phi}
\newcommand{\templatefunc}[1]{{\templatefuncsymbol(#1)}}
\newcommand{\simpleof}[1]{{#1^{*}}}
\newcommand{\rvaluations}{{\valuationspace_\rasiowa}}

\section{The Rasiowa Model}
Given any logic $\logic$ which is an \techterm{axiomatic extension} of $\delogic$, a model for $\logic$ is constructed
which I call the $\deftechterm{Rasiowa model}\ \rasiowa$ of $\logic$. 
Let $\sig$ be the signature of $\logic$, and let $\terms$ be the set of terms with respect to $\sig$.
On $\terms$, define the relation $\termeq$ by
\[
    \termeqapp s t \quad\text{iff}\quad \logic \derives \opequalsapp s t.
\]
Clearly, $\termeq$ is an equivalence relation. For example, $\termeq$ is reflexive because of Axiom
$\textsc{Equality}_1$ of $\delogic$. Let $\classof t$ be the equivalence class of $t \in \terms$
with respect to $\termeq$, and let $\termclasses$ be the set of all such equivalence classes.
Then the Rasiowa model
\[\rasiowa = (\algebra_\rasiowa, \classof \true, \rvaluations)\]
has $\termclasses$ as the carrier of $\algebra_\rasiowa$. It remains to define the operations
of the $\sig$-algebra $\algebra_\rasiowa$, and to define the valuation space $\rvaluations$.
Each $n$-ary template $T = [x_1 \ldots x_n.\, t]$ can be viewed as a function 
$\templatefunc T : \termclasses^n \rightarrow \termclasses$
which maps $(\classof{r_1}, \ldots, \classof{r_n})$ to $\classof{\appsubst \sigma t}$, where 
$\sigma = \{\appsubst {x_1} {r_1}, \ldots, \appsubst {x_n} {r_n}\}$ is the substitution that 
replaces $x_i$ by $r_i$. To see that $\templatefunc T$ is well-defined, assume $\termeqapp {r_1} {s_1}, \ldots, \termeqapp {r_n} {s_n}$
and note that 
$
\opequalsapp {x_1} {y_1} \opimplies \cdots \opimplies \opequalsapp {x_n} {y_n} 
\opimplies \opequalsapp {A[x_1, \ldots, x_n]} {A[y_1, \ldots, y_n]}
$ 
is a theorem of $\logic$. Instantiating $A$ with $[x_1 \ldots x_n.\, t]$, $x_i$ with $r_i$, and $y_i$ with $s_i$ then proves
together with $n$ applications of Modus Ponens that
$\termeqapp {\appsubst {\{\appsubst {x_1} {r_1}, \ldots, \appsubst {x_n} {r_n}\}} t} 
{\appsubst {\{\appsubst {x_1} {s_1}, \ldots, \appsubst {x_n} {s_n}\}} t}$. 
For $n = 0$,
the template $T$ is just a term $t$, and 
$\templatefunc T = \templatefunc t = \classof {\appsubst {\{\}} t}  = \classof t$.
This motivates extending $\termeq$ from terms to templates, such that $\termeqapp S T$ iff 
$\templatefunc S = \templatefunc T$. 

Now let $a \in \sig$ be an abstraction of shape $[p_1, \ldots, p_n]$. 
The operator $o$ associated with $a$ is constructed by defining $o(f_1, \ldots, f_n)$ as follows. 
If for each operation $f_i$, there is a template
$G_i$ such that $\templatefunc {G_i} = f_i$, then we choose the $G_i$ such that
$G_i = [x_{\alpha_1^i} \ldots x_{\alpha_{\abs {p_i}}^i}.\, g_i]$,
where $p_i = \{\alpha_1^i < \cdots < \alpha_{\abs {p_i}}^i\}$.
Here $x_1, \ldots, x_m$ are any $m$ distinct variables, and $m$ is the valence of $a$. Then 
$o(f_1, \ldots, f_n) = \classof{(a\, x_1 \ldots x_m.\, g_1\ldots g_n)}$.
This is again well-defined, because if we happen to choose different 
$G_i' = [y_{\alpha_1^i} \ldots y_{\alpha_{\abs {p_i}}^i}.\, g'_i]$, then nevertheless
$\termeqapp {G_i} {G'_i}$, and therefore \linebreak
$\termeqapp {(a\, x_1 \ldots x_m.\, g_1\ldots g_n)} {(a\, y_1 \ldots y_m.\, g'_1\ldots g'_n)}$.
Otherwise, if one of the $f_i$ is not in the image of $\templatefuncsymbol$, 
an arbitrary result is chosen, say $o(f_1, \ldots, f_n) = \classof {\true}$.

A valuation $\nu$ into $\algebra_\rasiowa$ is called \deftechterm{simple} if for each
variable $x$ and every arity $n$, there is a template $G$ such that $\nu(x, n) = \templatefunc G$.
Another way of describing simple valuations is via substitutions defined everywhere.
For such a substitution $\sigma$, define the valuation $\simpleof \sigma$ by 
$\simpleof \sigma(x, n) = \templatefunc {\sigma(x, n)}$. The valuation $\simpleof \sigma$ is 
obviously simple. Similarly obvious is that every simple valuation $\nu$ can be represented
as $\nu = \simpleof \sigma$ for some substitution $\sigma$ defined everywhere. 
It is straightforward to prove the following important property for simple valuations:
$\valueof {\simpleof \sigma} t = \classof{\appsubst \sigma t}$.

\newcommand{\canonical}{\kappa}
The space $\rvaluations$ then consists of those valuations which are simple. It is indeed
a \techterm{valuation space}. Firstly, $\rvaluations$ is inhabited: Let $\canonical$
be the \deftechterm{canonical substitution} defined by 
$\canonical(x, n) = [y_1 \ldots y_n.\, x[y_1, \ldots, y_n]]$. 
Then $\simpleof \canonical \in \rvaluations$. Secondly, assume
$\nu \in \rvaluations$, $u \in \termclasses$, and let $x$ be a variable.
Then there are $\sigma$ and $t$ such that $\nu = \simpleof \sigma$ and $u = \classof t$. Thus,
$\nu[x \coloneqq u] = \simpleof \sigma[x \coloneqq \classof t] = \simpleof {(\sigma[x \coloneqq t])} \in \rvaluations$.
Thirdly, assume $\nu \in \rvaluations$, and let $\theta$ be a substitution with domain $D$. Then there is 
$\sigma$ such that $\nu = \simpleof \sigma$. For $(x, n) \notin D$, 
$\nu_{\theta}(x, n) = \nu(x, n) = \templatefunc {\sigma(x, n)}$. For $(x, n) \in D$, 
$\nu_{\theta}(x, n) = \templatefunc {\appsubst \sigma {\theta(x,n)}}$. 
Together, this yields $\nu_\theta \in \rvaluations$.

To show that $\rasiowa$ is indeed a model of $\logic$, we need to confirm that all inference rules of 
$\logic$ are true in $\rasiowa$.

To confirm \textsc{Modus Ponens}, assume that both $\opimpliesapp A B$ and $A$ are true in $\rasiowa$
for valuation $\nu =\simpleof \sigma \in \rvaluations$. This means 
$\classof \true = \valueof \nu {\opimpliesapp A B} = \valueof {\simpleof \sigma} {\opimpliesapp A B}  = 
\classof{\appsubst \sigma {(\opimpliesapp A B)}} = \classof {\opimpliesapp {\appsubst \sigma A} {\appsubst \sigma B}}$,
and similarly $\classof \true = \classof {\appsubst \sigma A}$. 
This means that both $\opequalsapp \true {(\opimpliesapp {\appsubst \sigma A} {\appsubst \sigma B})}$
and $\opequalsapp \true {\appsubst \sigma A}$ are theorems in $\logic$, and furthermore 
we can deduce the theorem $\opequalsapp \true {(\opimpliesapp \true {\appsubst \sigma B})}$, then
$\opimpliesapp \true {\appsubst \sigma B}$, and then $\appsubst \sigma B$. Using Axiom $\textsc{Truth}_2$
we obtain $\opequalsapp {\appsubst \sigma B} \true$, which means that $\valueof \nu B = \classof\true$.

To confirm \textsc{Universal Intro}, assume that $[x.\,P[x]]$ is true in $\rasiowa$ for 
valuation $\nu =\simpleof \sigma \in \rvaluations$. This means that the operation
$u \mapsto \valueof {\nu[x \coloneqq u]} {P[x]}$ is equal to $\classof \true$ everywhere. Therefore,
for all $t \in \terms$ we have $\classof \true = \valueof {\nu[x \coloneqq \classof t]} {P[x]} =
\valueof {(\simpleof \sigma)[x \coloneqq \classof t]} {P[x]} =
\valueof {\simpleof {(\sigma[x \coloneqq t])}} {P[x]} =
\classof{\appsubst {\sigma[x \coloneqq t]} {P[x]}} = \classof{\sigma(P, 1)(t)}$, which means
that in $\logic$ the theorem $\opequalsapp \true \sigma(P, 1)(t)$ is deducible for any term $t$.
Thus, \linebreak $\valueof \nu {\forall x.\, P[x]} = \valueof {\simpleof \sigma} {\forall x.\, P[x]} =
\classof {\appsubst \sigma {(\forall x.\, P[x])}} = \classof {\forall x.\, \sigma(P, 1)(x)} = 
\classof {\forall x.\, \true}$. As $\forall x.\, \true$ is a theorem of $\delogic$, this means
that $\valueof \nu {\forall x.\, P[x]} = \classof {\forall x.\, \true} = \classof \true$.

To confirm that every \techterm{axiom} $t$ of $\logic$ is true in $\rasiowa$ for every 
$\nu = \simpleof \sigma \in \rvaluations$, note that $\valueof \nu t = \valueof {\simpleof \sigma} t
= \classof {\appsubst \sigma t}$. But since $t$ is an axiom, $\logic \derives t$, and therefore
also $\logic \derives \appsubst \sigma t$. This implies 
$\valueof \nu t = \classof {\appsubst \sigma t} = \classof \true$.

Therefore, $\rasiowa$ is a model of $\logic$.

\section{Completeness and Consistency}
An abstraction logic $\logic$ is \deftechterm{complete} if every valid term is also a theorem, i.e.\ if
\[
\logic \valid t \quad \text{implies}\quad \logic \derives t\quad\text{for every term $t$.}
\]
If $\logic$ is an axiomatic extension of $\delogic$, then it is complete. To see this, construct
the Rasiowa model of $\logic$. If $t$ is a valid term, then $t$ must be true in every
model of $\logic$, and therefore also in the Rasiowa model. Thus, 
$\valueof \nu t = \classof \true$ for any valuation $\nu \in \rvaluations$. 
In particular, using the canonical substitution $\canonical$,
$\classof \true = \valueof {\simpleof \canonical} t = \classof{\appsubst {\canonical} t} = \classof{t}$.
But this means that $\opequalsapp \true t$ is a theorem of $\logic$, and thus $\logic \derives t$.

An abstraction logic $\logic$ is called \deftechterm{inconsistent} if all terms are also theorems, 
and consequently \deftechterm{consistent} if there is at least one term which is not a theorem.
If $\logic$ is an axiomatic extension of $\delogic$, then inconsistency of $\logic$ is equivalent to 
$\logic \derives \forall x.\, x$.
This is easy to see. Assume $\forall x.\,x$ is a theorem in $\logic$. 
Substituting $[x.\, x]$ for $A$ in Axiom~$\textsc{Universal}_1$ and applying Modus Ponens, 
it follows that $x$ is a theorem. Substituting any term $t$ for $x$ shows that $t$ is a theorem. 

If an abstraction logic $\logic$ is inconsistent, then all models of $\logic$ are degenerate. 
To see this, assume $\logic$ is inconsistent. It follows that $x$ is a theorem for any variable $x$. 
That means that $x$ is valid in any model $(\algebra, \true, \valuationspace)$ of $\logic$. 
In particular, for any valuation $\nu$ in $\valuationspace$ we have $\valueof \nu x = \true$. 
There is such $\nu$ because $\valuationspace$ is non-empty. For any value $u \in \carrier \algebra$,
the valuation $\nu[x \coloneqq u]$ is in $\valuationspace$ as well,
and therefore $u = \valueof {\nu[x \coloneqq u]} x = \true$. That means $\carrier \algebra = \{\true\}$.
The logics $\logic_\peanoalgebra$, $\logic_\setsalgebra$ and $\logic_\funsetsalgebra$ are therefore
all consistent, because they have non-degenerate models.

What about the other direction? If every model of an abstraction logic $\logic$ is degenerate, then
$\valueof \nu t = \true$ holds for any term $t$ with respect to any valuation $\nu$, and therefore $\logic \valid t$.
If $\logic$ is complete, this implies $\logic \derives t$, making $\logic$ inconsistent. 

For complete logics $\logic$, such as all logics axiomatically extending $\delogic$, 
inconsistency of $\logic$ is therefore equivalent to all models of $\logic$ being degenerate. 

\section{Related and Future Work}

Given my claim that AL is the right logic, the related work section should probably be a paper on its own. 
This is of course not possible here. Rather, I consider connecting AL to established work in logic
to be future work. For example, it seems straightforward to represent type theory and intuitionistic logic
as abstraction logics. What does this mean for type theory and intuitionistic logic? How do 
Henkin models for simple type theory relate to AL models when representing simple type theory as an AL, 
and/or when encoding AL via simple type theory? And so on.
I will therefore just touch briefly upon two bodies of work immediately related to AL.\ 

Abstraction logic is a generalization of Rasiowa's work on \techterm{algebraic semantics} for 
non-standard propositional logics~\cite{rasiowa}. 
Rasiowa chose to generalize her approach to predicate logic by following Mostowski and interpreting 
quantifiers as least upper bounds and greatest lower bounds. To me, that does not seem to be the proper way 
to generalize Rasiowa's approach beyond propositional logic. Instead, I believe abstraction logic is.

A major inspiration for AL (and Practal) is the Isabelle logical framework~\cite{isabelle}. 
Its meta-logic $\mathcal{M}$~\cite{isabelle-foundation} is based on 
intuitionistic higher-order logic, which is based on the $\lambda$-calculus.
Like most other logical frameworks, Isabelle is therefore also a case of 
\emph{Just Use a Lambda\texttrademark}.
To justify that an object logic built on top of $\mathcal{M}$ is sound and complete,
one has to make an argument specifically for each such object logic.
In contrast to this, \emph{every} abstraction logic is sound with respect to a simple semantics, 
and if it is axiomatically extending $\delogic$, it is also guaranteed to be complete.
Still, AL is also just algebra, and this seems to bode well for practical applications,
suggesting AL as a foundation not only for proof assistants, 
but also for computer algebra systems, automated theorem proving, and any other application
in need of a general formal language based on a simple semantics.

\section{Acknowledgements}
Norbert Schirmer commented on an early version of this paper, 
and asked revealing questions about AL that helped me in shaping the narrative and structure of this paper.


\bibliographystyle{amsalpha}


\end{document}